\documentclass[runningheads,a4paper,oribibl,12pt]{llncs}
\usepackage{graphicx}
\usepackage{natbib}
\usepackage{makeidx}
\RequirePackage[colorlinks,citecolor=blue,urlcolor=blue]{hyperref}
\usepackage{mathptmx}
\usepackage{dcolumn}
\usepackage{bm}
\RequirePackage{mathrsfs,amssymb}

\begin{document}
\frontmatter
\pagestyle{headings}
\title{Tax Compliance and Public Goods Provision\\  - An Agent-based
Econophysics Approach -}

\titlerunning{ }
\titlerunning{ }
\author{Sascha Hokamp\inst{1} \and G\"otz Seibold\inst2}
\institute{Center for Earth System Research and Sustainability, 
         University of Hamburg, Grindelberg 7, 20144 Hamburg, Germany \\ 
         \email{sascha.hokamp@zmaw.de}
         \and Institute of Physics, BTU Cottbus -- Senftenberg, P.O.Box 101344,
         03013 Cottbus, Germany \\ \email{goetz@physik.tu-cottbus.de}}

\maketitle

\begin{abstract}
We calculate the dynamics of tax evasion within a multi-agent
econophysics model which is adopted from the theory of magnetism
and previously has been shown to capture the main characteristics
from agent-based based models which build on the standard Allingham and Sandmo
approach.
In particular, we implement a feedback of public goods provision on the 
decision-making of selfish agents which aim to pursue their self interest. 
Our results imply that such a feedback 
enhances the moral
attitude of selfish agents thus reducing the percentage of tax evasion. 
Two parameters govern the
behavior of selfish agents, (i) the rate of adaption to changes in
public goods provision and (ii) the threshold of perception of public goods
provision. Furtheron we analyze the tax evasion dynamics for different agent compositions and under the feedback of public goods provision.
We conclude that policymakers may enhance tax compliance behavior via the 
threshold of perception by means of targeted public relations.\\

\vspace*{1cm}

{\bf JEL classification}: C15, H16, H30\\
\keywords{tax compliance; econophysics; multi-agent model}
\end{abstract}

\pagebreak

\section{Introduction}\label{intro}
Theoretical approaches to account for tax compliance are often based on the 
seminal work of \citet{alling72} 
which incorporates potential penalties, tax rates and audit probabilities as basic 
parameters in order to evaluate the expected utility of taxpayers. 
In the most basic version of this neoclassical standard scheme the behavior of
{\it all} taxpayers enters via the {\it average} degree of risk aversion, 
related to the structure of the underlying utility function. 
However, the reasons and motivations for tax evasion in a
society are manifold and it would obviously be desirable to relate the 
behavior of individuals, embedded in their social network, to
global parameters such as risk aversion.
In some sense this hierarchy of modelling has a close match in physics
where phenomenological laws of thermodynamics can be derived within
the concepts of statistical mechanics which are applied to the individual
constituents of a macroscopic system. For example, thermodynamic parameters
such as the temperature of an ideal gas can be traced back to the average 
energy which is distributed among the various degrees of freedom of the 
individual gas molecules.

Within the economics domain, in particular tax evasion, the approach
by \citet{alling72} would correspond to the phenomenological level
of description. On the other hand so-called agent-based models have been
set up as a comparatively new tool for analyzing 
tax compliance issues on a more individual (i.e. 'microscopic') level.
An essential feature of any agent-based model 
is the direct interaction between agents, which is 
combined with some process that allows for changes in individual 
behavior patterns. 

According to \citet{hopi10} and \citet{pick12} agent-based tax evasion models 
may be categorized into an economics and econophysics domain. 
The latter has been initiated by 
\Citet{lz08,zak08,zak09} and \citet{lima10} by analyzing a suitable
modification of the Ising model \citep{ising25}, originally known from 
the theory of magnetism.
In contrast, if the interaction process is driven by parameter changes that 
induce behavioral 
changes via utility functions and (or) by stochastic processes that do 
not have physical roots, these models belong to the economics domain. 
Examples include \cite{mp00,dav03,bloom04,bloom08,antun07,koro07,szabo09,szabo10,hopi10,meder10,nordblom12,andr13,ho2014,pelli13} of which some are summarized by \cite{bloom06}, \cite{pickprinz}, and \citet{pick12}. 

In previous work  \cite{seibpick12, pick12} and more recently \citet{hoseib} 
have extended the Ising-based econophysics approach
to tax evasion toward the implementation of different agent types.
This theory is able to reproduce results from agent-based economics
models \citep{hopi10} and therefore should be appropriate for a quantitative
analysis of tax compliance. Moreover, the econophysics route to tax
evasion may provide the formal framework to construct a phenomenological
'global' theory starting from a microscopic 'agent-based'
description. However, before tackling this ambitious task it is
of course necessary to provide an econophysics description of tax
evasion which comprises the main ingredients inherent in contemporary
neoclassical approaches based on the work by \citet{alling72}.

In this regard, one crucial aspect which we aim to improve in the
present paper
concerns the time evolution of social norms within the network
of agents. In fact, all previous econophysics works mentioned above 
classified the individual agents by two
parameters, (i) a local field describing the moral attitude of agents
toward tax evasion and (ii) a local temperature which governs the susceptibility
of agents with regard to behavioral changes. Both parameters compete
with the interaction between agents which aims to conform the agent's behavior
within their social network.

However, these parameters were fixed and therefore behavioral changes
only occured as a result of the statistical evaluation of the dynamics
incorporating the aforementioned competition between interaction of agents
and their moral attitude and/or local temperature, respectively.

Here we aim to go beyond this static description by incorporating
social norm updating into the agent-based econophysics approach
to tax evasion. In particular, we allow for dynamical changes
in the moral attitude of agents due to public goods provision.
Within standard agent-based models this issue has been
studied by \cite{mp00,antun07,szabo09,szabo10,meder10,ho2014,pelli13} 
with partially contradictory results.
For example, \citet{szabo09,szabo10} showed that increasing the level of
governmental services (e.g. health care) leads to less tax evasion. Note 
that the authors consider more than 20 employment types and four agent types, 
namely (i) elected administration, (ii) tax authority, (iii) workers, 
and (iv) entrepreneurs. In contrast, \cite{ho2014} finds the counterintuitive
result that income tax compliance may decrease with raising marginal per capita
returns. The latter analysis is based on a model with back auditing
and four agent types which are also implemented in the present paper
(cf. Sec. \ref{sec2}).
 
The paper is organized as follows. In order to put our investigations
into the appropriate economic context
we present in Sec. \ref{secpg} a brief literature review 
regarding public goods provision and tax compliance. In Sec. \ref{sec2} 
we outline the basic ingredients of our econophysics
model [cf. \cite{seibpick12,hoseib,pick12}] and exemplify the approach for a 
society with homogeneous agents which sets the stage for the following
discussion. The procedure how the provision of public goods influences on
selfish agents within our model is presented in 
Sec. \ref{sec3} and corresponding
results are shown in Sec. \ref{secres}. We finally conclude our 
discussion in Sec. \ref{sec4}. 

\section{Public Goods Provision and Tax Compliance: A Literature Review}
\label{secpg}
The modern root of public goods theory dates back to the seminal work of 
\citet{SAMU1} introducing a condition for Pareto-optimal allocations: the 
sum of individual's marginal rate of substitution has to equal the marginal rate of 
transformation. \Citet{PICK1,PICK3} surveys modern public goods theory and the relevant literature therein, e.g. \citet{MUSG1,SAMU1,SAMU2,MUSG3,SANO}. Among other things, he observes a large diversity in defining and using the terminus 'public good' that is to some extent conflicting or even contradicting. In our paper we adopt the consumption-act-approach
of \citet{PICK1}, i.e. properties of goods are totally atomized into acts of consumption at any point in time and then 'rival' and 'non-rival' consumption acts are used to 
determine private and public goods characteristics, respectively. Note that the 
terminus 'non-rival in consumption' was introduced to the literature 
by \citet[pg. 126]{MUSG2} meaning '[\dots] that the same physical output 
(the fruits of the same factor input) is enjoyed by both [individuals,] A 
and B. This does not mean that the same subjective benefit must be derived, 
or even that precisely the same product quality is available to both.' 
However, with respect to income tax evasion and public goods provision we briefly review below theoretical works of \citet{COGO,FALK1,FALK2, falki,COWE} and \citet{BORD}, as well as experiments by \citet{alm1,alm2,mittone06,ALM2010} and \citet{bapick11}. For details see \citet{ho2013}.

\Citet{COGO} postulate a two dimensional 
continuum of goods; on the one hand bounded by rivalness and non-rivalness 
and on the other hand limited by 'excludability' and 
'non-excludability'. Note that in the course of our paper excludability in consumption is assumed if and only if 
options are feasible and enforceable to ban individuals from consumption of these 
goods. \Citet{COGO} examine the influence of public goods provision on individual's and average population's tax evasion behavior. The authors investigate (i) large populations, where a single 
individual decision to commit tax evasion has no observable impact on public goods provision and (ii) decreasing absolute risk 
aversion when increasing income. In addition, they assume under-provision 
(over-provision) of public goods to deduce their 
counterintuitive key findings: raising a flat tax ceteris paribus increases 
(decreases) tax evasion.

\Citet{FALK1} adds the notion of tax as a price for public services to the 
basic flat tax setting of \citet{alling72}. Note that these governmental provided public goods are 
non-rival -- i.e. each agent may consume the same amount -- but individual's 
utility of consumption may differ. \Citet{FALK1} then 
examines equity of government-taxpayer relationships under presence of tax 
non-compliance and public goods provision. He finds that tax evasion ceteris paribus is reduced if four assumptions are 
simultaneously fulfilled: (i)~public goods provision only depends 
on expected tax revenues, (ii)~behaviors of other agents do not change 
(the 'Nash-assumption', cf.~ibid., pg. 390), (iii)~tax enforcement 
variables have to allow rational for a positive tax evasion, and (iv)~an 
additive utility function prevails. Eventually, his main insight is '[\dots] that the equity 
argument is [rather] an ex post rationalization of otherwise 
motivated tax evasion' than an ex ante driving force \citep[see][pp. 392-393]{FALK1}. 

\Citet{FALK2} addresses a reverse question related to \citet{COGO}, that 
is: How might tax evasion influence the optimal level of public goods 
provision? Using additive and 
strictly concave utility functions \citet{FALK2} shows that the impact of 
flat tax evasion on the optimal supply of public goods depends on the 
magnitude of decreasing individual's absolute risk aversion. To analyze inequity \citet{COWE} enlarges the publicly-supplied-goods-approach of \citet{COGO} 
while maintaining their notion of public goods (cf.~above). As pointed out by \citet[][pg. 541]{COWE},
 the essential problem of economic, psychological and social effects -- 
which simultaneously influence individual's flat tax evasion behavior and 
public goods provision -- concentrates to one single crucial question:
 '[c]an the person affect inequity directly by his own actions?'.

\Citet{BORD} extends the setting of \citet{COGO} via introducing 
taxpayers' heterogeneity that refers to income, magnitude of private 
consumption and audit probability. In particular, \Citet[][pg. 359]{BORD} examines '[\dots] the fairness of the fiscal 
system [\dots]' depending on tax structure, public goods provision and 
subjects' beliefs or individuals' estimates of other peoples' tax evasion. 
He then finds that poor taxpayers ceteris paribus evade more (less) in absolute terms 
than rich subjects, if the whole population is faced with low (high) tax 
rates. To allow for a comparison with \citet{FALK1} and \citet{COWE} the author drops the assumption of agents' heterogeneity and 
derives Pareto-optimal provision levels of public goods. Then~--~within his 
homogeneous fairness-approach assuming under-provision (over-provision) 
of public goods -- \citet{BORD} shows that raising the tax rate leads to 
zero (positive) tax evasion; where positive tax evasion ceteris paribus increases
 (decreases) with low (high) tax rates. \Citet{falki} takes an economic and a psychological point of view on flat 
tax evasion and equity that both base on \citet{FALK1,FALK2} and \citet{COWE}.
 Yet, \citet{falki} explains counterintuitive results of \citet{COWE} through a decline of 
absolute risk aversion while increasing equity.

\Citet{alm1} present a tax compliance experiment and find that providing a surplus from tax payments -- i.e. a public good -- may work to increase tax compliance. \Citet{alm2} provide an experimental contribution to taxpayer's mystery, 
that is, '[w]hy do people pay taxes?' Regarding this voluntary contribution puzzle \citet{alm2} argue 
that (i)~taxpayers may value public goods provision, (ii)~individuals probably
 err on the real extent of tax enforcement variables -- in particular, they
 might overweight audit probabilities -- and (iii)~subjects may hold an extreme risk aversion. Further, the authors obtain that flat tax evasion even occurs if expected returns of a tax gamble are negative. \Citet{mittone06} conducts a dynamic tax evasion experiment under risk and uncertainty. Among other things, the author finds that '[\dots] the production of a public good reduces tax evasion but to a lesser extent than tax yield distribution' \citep[][pg. 830]{mittone06}. \citet{alm2} conclude that new theories are quite necessary to explain taxpayers' responses 
frequently observed in laboratory experiments with respect to taxes, evasion, 
public goods provision, and the like. An interdisciplinary approach, e.g. behavioral economics, may fill in
 this gap. See economist (and chemist) \citet[][pg. 636]{ALM2010} for
 a history of the terminus 'behavioral economics', meaning '[\dots] an
 approach that uses methods and evidence from other social sciences 
(especially psychology) to inform the analysis of individual and group
 decision making'. For a psychologist's point of view see \citet{KIRC}. 

\Citet{ALM2010} briefly surveys behavioral (public) economics; an
 interdisciplinary research field that deviates from several standard 
neoclassical assumptions, for instance, subjects may not always act rational
 and selfish. In particular, the author distinguishes three experimental branches that
 examine (i)~'public goods', (ii)~'tax compliance' and
 (iii)~'behavioral responses to taxes' (cf.~ibid., pg. 635). Furthermore, \Citet[][pg. 650]{ALM2010} presents a brief
 overview of open tasks and promising ideas for future experiments in the 
domain of behavioral public economics, which concern (i)~'intertemporal 
decisions', (ii)~'social insurance programs' and (iii)~'behavior under 
uncertainty'. Finally, note that the author quotes the agent-based approach
 of \citet{BLOO4} to illustrate a fruitful combination of laboratory
 experiments to computational simulations for investigating the evolution of
 tax compliance dynamics over time.

\Citet{bapick11} conduct a flat tax compliance experiment that examines the 
relationship of tax evasion and positive incentives under public goods
 provision and lottery effects. Note that the seminal work of \citet{FAWA} has added welfare improving positive rewards for audited individuals to the standard model of tax evasion, that is \citet{alling72}; i.e. taxpayers get a monetary payback for their voluntarily paid taxes regardless of their actual extent of tax cheating. On contrary, \citet{bapick11} employ
 non-rival public goods and install an excludable flat tax gamble, where only
 fully compliant taxpayers can get the extra profit. In each round of the
 laboratory experiment the public good as well as the permissible extra gain
 is financed by voluntary contributions~--~i.e. taxes~--~forced reimbursements
 and paid penalties, so that each individual influences marginally the extent
 of public goods provision. \citet{bapick11} find experimental evidence for
 gender effects: (i)~females respond stronger to an increase of
 penalties than males, (ii)~females evade less than males, and (iii)~males
 react stronger than females to positive lottery rewards -- ceteris paribus yielding higher
 tax compliance rates.

Finally, to sum up reviewed neoclassical theoretical and experimental 
applications present a wide range of conclusions, effects, findings and
 results that may appear to some extent ambiguous, counterintuitive, or 
actually conflicting and contradicting with respect to income tax evasion, 
public goods provision and their inter dependencies. Analogous discrepancies
also appear in agent-based models as already outlined in Sec. \ref{intro}.
In the next section we proceed by introducing our econophysics agent-based
model for which we implement the influence of public goods in Sec. \ref{sec3}.

\section{The Agent-Based Econophysics Approach}\label{sec2}
Our considerations are based on the Ising model hamiltonian
\begin{equation}\label{eq:isi}
H=-J \sum_{\langle ij\rangle}S_i S_j - \sum_i B_i S_i
\end{equation}
where $J$ describes the coupling of Ising variables (spins) $S_i=+1$,$-1$ 
between adjacent lattice sites denoted by $\langle ij\rangle$. 
In the present context $S_i=+1 (-1)$ is interpreted as a
compliant (non-compliant) taxpayer.
Calculations have been done for 
a two-dimensional square lattice with dimension $1,000 \times 1,000$
and periodic boundary conditions, i.e. a torus structure.
We note that the results are not sensitive to the specific lattice geometry. 
Within a similar econophysics model this issue has been 
analyzed by \cite{zak09,zak08} and \cite{lz08} 
who consider, in addition, 
alternative lattice structures like the scale-free Barab\'{a}si-Albert 
network or the Voronoi-Delaunay network. In addition \cite{lima10} considers 
Erd\"os-R\'{e}nyi random graphs and finds that the results for these 
alternative lattices do not differ significantly from 
those obtained with a square lattice.

Eq. (\ref{eq:isi}) contains also
the coupling of the spins to a local magnetic field $B_i$
which can be associated with the morale attitude of the agents 
and corresponds to the parameter
$\gamma_i$ in the theory of \citet{nordblom12} which have
investigated social norm updating within an agent-based model.
In addition, our model contains a 
local temperature $T_i$ which measures the susceptibility
of agents to external perturbations (either influence of neighbors
or magnetic field).  We then use the heat-bath 
algorithm [cf. \cite{krauth}] in order to evaluate statistical 
averages of the model. The corresponding program
code is written in FORTRAN and is available upon request.
The probability for a spin at lattice site $i$ to take the values 
$S_i=\pm 1$ is given by 
\begin{equation}\label{eq:prob} 
p_i(S_i)=\frac{1}{1+\exp\{-[E(-S_i)-E(S_i)]/T_i\}} 
\end{equation}
and $E(-S_i)-E(S_i)$ is the energy change for a spin-flip at site $i$.
Upon picking a random number $0 \le r \le 1$ the spin takes the value
$S_i=1$ when $r < p_i(S_i=1)$ and $S_i=-1$ otherwise.
One time step then corresponds to a complete sweep through the lattice.

Following \cite{dav03} and \cite{hopi10} we consider societies which are 
composed of the following four types of agents:
(i) \textit{selfish a-type agents}, which take advantage from 
non-compliance ($S_i=-1$) and, thus, are characterized 
by $B_i/T_i < 0$ and $|B_i| > J$; (ii) \textit{copying b-type agents}, 
which conform to the 
norm of their social network and thus copy the behavior with respect
to tax evasion from their neighborhood. This can be modeled
by $B_i << J$ and $J_i/T_i \gtrsim 1$; 
(iii) \textit{ethical c-type agents}, which are practically
always compliant  ($S_i=+1$) and thus are parametrized by 
$B_i/T_i > 0$ and $|B_i| > J$; 
(iv) \textit{random d-type agents}, which act by chance, within a 
certain range, due to some confusion caused by tax law complexity.
We implement this behavior by $B_i<<J$ and $J/T_i <<1$. 
The parameters distinguishing the different agent types are
taken from \citet{seibpick12} and \citet{pick12}. 

In analogy to \citet{zak08,zak09,lima10} and \citet{pick12} 
we further
implement the probability $p_a$ of an audit. 
If tax evasion is detected the agent is forced to stay compliant 
over $h$ time steps (penalty period). 
Such a procedure has also been
implemented in a randomized variant in \citet{lz08}.
We also note that it is possible to incorporate 
lapse of time effects, i.e. the situation where a
detected agent is also screened over several years in the
past by the tax authorities (i.e. backaudit). This variant has been 
studied within an agent-based econophysics tax compliance model in \citet{seibpick12}.

\begin{figure}[htb]
\begin{center}
\includegraphics[width=12cm,clip=true]{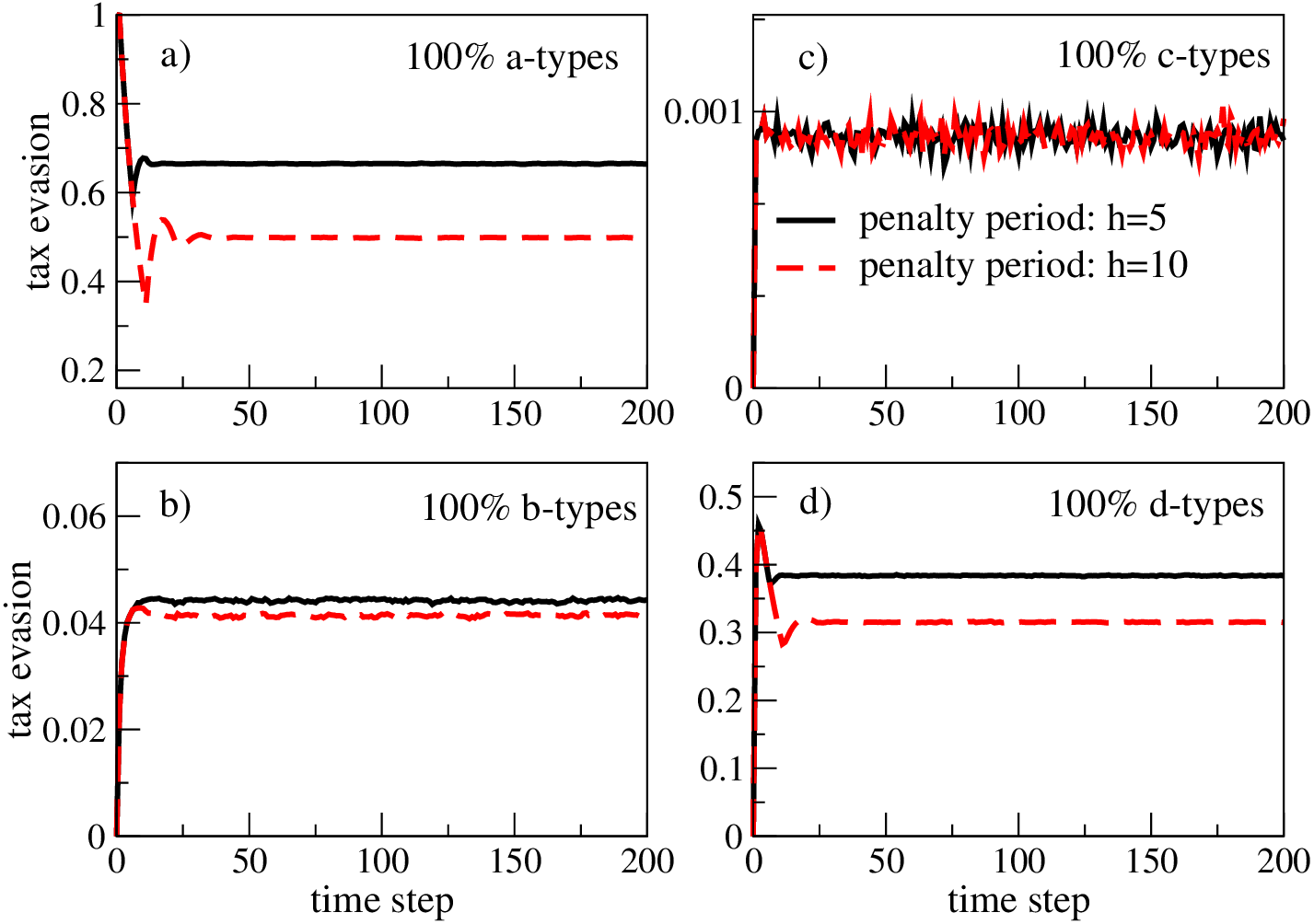}
\end{center}
\caption{Time evolution of tax evasion corresponding to the fraction 
of non-compliant taxpayers $p_{non-cp}$. The panels report results
for a society consisting of $100\%$ a-type (panel a), 
$100\%$ b-type (panel b), $100\%$ c-type (panel c), 
and $100\%$ d-type (panel d) agents.
Solid and black lines display the tax evasion dynamics for penalty  period 
$h=5$ whereas dashed and red lines are for penalty period $h=10$. 
Audit probability is $p_a=10\%$ in each case.}
\label{fig0}
\end{figure}

Since in the following we implement the influence of public goods provision
which affects selfish a-types only, it is instructive to  consider the case 
of a society with all agents being of the same type. 
The resulting percentage of tax evasion
as a function of time is shown in Fig. \ref{fig0}
for two penalty periods ($h=5,10$).

The first case of endogenous non-compliant selfish agents  is 
shown in Fig. \ref{fig0}a where at time step zero all agents
are set to non-compliance, i.e. the fraction of non-compliant taxpayers 
is $p_{non-cp}=1$. 
Due to the enforcement mechanism,  $p_{non-cp}$ is significantly
reduced because 
at each time step a certain percentage of the remaining non-compliant 
agents are forced to become compliant. Before reaching a stationary
value small oscillations are observed since
after  $h$ time steps the first detected 
agents can become non-compliant again. 
Fig. \ref{fig0}b reports the result for copying b-type agents.
As initial condition all agents are set to 'compliant'. Since b-types 
tend to copy the behavior of their social network only few of them change
their behavior and the equilibrium value for tax evasion approaches 
a rather small value
between $4\%$ and $5\%$. It should be noted that the equilibrium
value is independent of the initial condition. If we would have set
all agents to non-compliant at time step zero, the audits would have
reduced the percentage of tax evasion to the same equilibrium value than shown
in  Fig. \ref{fig0}b.

The time evolution for ethical c-type agents is reported in
Fig. \ref{fig0}c. Here the initial fraction of non-compliant taxpayers
is set to $p_{non-cp} =0$ and there is only a very 
small probability that one of the agents becomes non-compliant.  
Since ethical agents avoid tax evasion the results are 
also almost independent of the audit probability. 
Hence, any positive audit probability would be inefficient in this case. 

Finally, Fig. \ref{fig0}d shows the percentage of tax evasion
for random d-type agents. Since these agents act by chance their probability
for changing their behavior from compliant to non-compliant or vice versa
would be of the order of $\sim 50\%$ 
without any audit.
The enforcement mechanism then leads to a further 
reduction for the equilibrium value of $p_{non-cp} =0$ 
where larger penalty periods
naturally lead to a stronger reduction.

\section{The Influence of Public Goods Provision}\label{sec3}
In our previous considerations the local fields for the ethical (c-type)
and selfish (a-type) agents where initially fixed and independent
of the time evolution. For ethical agents this means a
deep seated moral attitude inculcated by education, religious
convictions [cf. \cite{heinemann12}] etc.
However, selfish agents are expected to maximize their personal benefit
by deliberating about whether their profit from public goods is supported,
independent or hindered by the tendency of tax evasion.
We therefore set up the following instructions for selfish
a-types: 

(a) Compare the provision of public goods between the current and
previous time step. Within our approach this provision  is
proportional to the fraction of compliant tax payers $p_{cp}$ 
(with $p_{cp}+p_{non-cp}=1$) and we therefore 
evaluate $\Delta p_{cp} = p_{cp}(t_n)-p_{cp}(t_{n-1})$. Moreover,
only a substantial improvement or decline $|\Delta p_{cp}|> \Delta p_{min}$ 
in the provision can be perceived by an individual, where 
the threshold of perception $\Delta p_{min}$ enters as an additional parameter.

(b) Each a-type agent compares its behavior between the current 
and previous time step, i.e. $\Delta S_i^a = S^a_i(t_n) - S^a_i(t_{n-1})$
where $i$ denotes the lattice site. The three possible values
$\Delta S_i^a=-2,0,2$ indicate a change from compliance to non-compliance, 
no behavioral change and a change from non-compliance to compliance, 
respectively.

(c) Change the attitude of a-types towards non-compliance (i.e. the
local field $B_i$ of a-types) by a step $\Delta B >0$ depending
on $\Delta p_{cp}$ and $\Delta S_i^a$. 

Instruction (c) corresponds to the following possibilities
which are in line with the analysis of \cite{falki} and
summarized in Tab. \ref{tab1}:

\begin{table}
\begin{center}
\begin{tabular}{|l|l|l|}\hline
Change of Public Goods Provision   & Change of a-type Behavior & Resulting Moral Attitude Change \\ \hline\hline
$|\Delta p_{cp}|<\Delta p_{min}$ & $\Delta S_i^a=-2, 0, 2$ & $0$ \\ \hline
$|\Delta p_{cp}|>\Delta p_{min}$ & $\Delta S_i^a=0$ & $0$ \\ \hline
$\Delta p_{cp}>\Delta p_{min}$ & $\Delta S_i^a=2$ & $+\Delta B_i$ \\ \hline
$\Delta p_{cp}<-\Delta p_{min}$ & $\Delta S_i^a=2$ & $-\Delta B_i$ \\ \hline
$\Delta p_{cp}>\Delta p_{min}$ & $\Delta S_i^a=-2$ & $-\Delta B_i$ \\ \hline
$\Delta p_{cp}<-\Delta p_{min}$ & $\Delta S_i^a=-2$ & $+\Delta B_i$ \\ \hline
\end{tabular}
\end{center} 
\caption{Summary of how changes in public goods provision
and behavioral change of a-types influence on the (im)moral attitude change of
these agents.}
\label{tab1} 
\end{table}

\begin{description}
\item [$|\Delta p_{cp}|<\Delta p_{min}$, $\Delta S_i^a=-2, 0, 2$:]
The provision of public goods does not change significantly,
i.e. it is below the threshold of perception $\Delta p_{min}$
of selfish agents. The
agent therefore keeps its moral attitude and the field
remains unaltered.
\item [$|\Delta p_{cp}|>\Delta p_{min}$, $\Delta S_i^a=0$:]
The provision of public goods has changed significantly
(i.e. it is above the threshold of perception $\Delta p_{min}$
of selfish agents) while the agent did not change its behavior.
It is not perceivable for the rational agent 
whether a change of its behavioral norm would lead to any
advantage. Therefore also in this case 
the agent tends to keep its moral attitude and the field
remains unaltered.
\item [$\Delta p_{cp}>\Delta p_{min}$, $\Delta S_i^a=2$:]
The agent has changed its behavior from non-compliant to compliant.
Simultaneously the provision of public goods has significantly
increased. The agent therefore recognizes a confirmation for its 
behavioral change 
and thus becomes less 'unethical'. This is implemented by a change
of its local field $B_i \rightarrow B_i+\Delta B$ which thus becomes
less negative. 
\item [$\Delta p_{cp}<-\Delta p_{min}$, $\Delta S_i^a=2$:]
The agent has changed its behavior from non-compliant to compliant but 
the provision of public goods has significantly
decreased. The agent therefore concludes that its behavioral change is not
honored and thus becomes more 'unethical'. The corresponding change
in the local field is $B_i \rightarrow B_i-\Delta B$.
\item [$\Delta p_{cp}>\Delta p_{min}$, $\Delta S_i^a=-2$:]
The agent has changed its behavior from compliant to non-compliant
while the provision of public goods has significantly
increased. The agent is therefore encouraged in its unethical
behavior which is implemented by a change in the local field
$B_i \rightarrow B_i-\Delta B$.
\item [$\Delta p_{cp}<-\Delta p_{min}$, $\Delta S_i^a=-2$:]
The agent correlates its change from compliant to non-compliant
behavior with the decrease of the provision of public goods suggesting
a more ethical behavior. Therefore the local field changes
as $B_i \rightarrow B_i+\Delta B$.
\end{description}

The new parameters entering our approach are thus
(i) $\Delta p_{min}$ (threshold of perception)
reflecting the lower bound in the
provision of public goods which can be perceived by an agent,
and (ii) the rate of adaption $\Delta B$ with which the a-type
agent adjusts its behavior to a change in the provision of public goods.
In our simulations we specify $\Delta p_{min}$ by a certain percentage
of compliant agents to which it is obviously proportional.
The parameter $\Delta B$ is chosen for each a-type agent $i$ as
a random number $0 < \Delta B_i < \Delta B_{max}$.

Table \ref{tab2} summarizes the parameters of our model.

\begin{table}
\begin{center}
\begin{tabular}{|l|l|l|}\hline
Parameter & Interpretation & Value \\ \hline\hline
Temperature T& Susceptibility to External Perturbations & 
$T=5$ for a- and c-types. \\  
& & $1 < T <3$ (b-types); $10 < T < 30$ (d-types)\\ \hline
Magnetic Field B & (Moral) Attitude towards& $B=0$ (b-,  d-types); $10 < B < 20$ (c-types)\\
& Tax Evasion & $B$ is self-consistently determined for a-types. \\ \hline
Audit Probability $p_a$ & & $p_a=0.1$ \\ \hline
Penalty Period $h$& & $h=5, 10$ time steps \\ \hline
$\Delta B_{max}$ & Maximum Rate of Adaption to & $\Delta B_{max}= 1 \dots 5$ 
\\ 
& Changes in Public Goods Provision & \\ \hline  
$\Delta p_{min}$ & Threshold of Perception for & $\Delta p_{min}=1\%$, $5\%$ \\
 
&  Changes in Public Goods Provision & \\ \hline  
\end{tabular}
\end{center} 
\caption{Compendium of parameters determining the present multi-agent model.
Note that temperature and magnetic field are measured in units of the
exchange coupling $J\equiv 1$.}
\label{tab2} 
\end{table}

\section{Results}\label{secres}

\begin{figure}[htb]
\begin{center}
\includegraphics[width=12cm,clip=true]{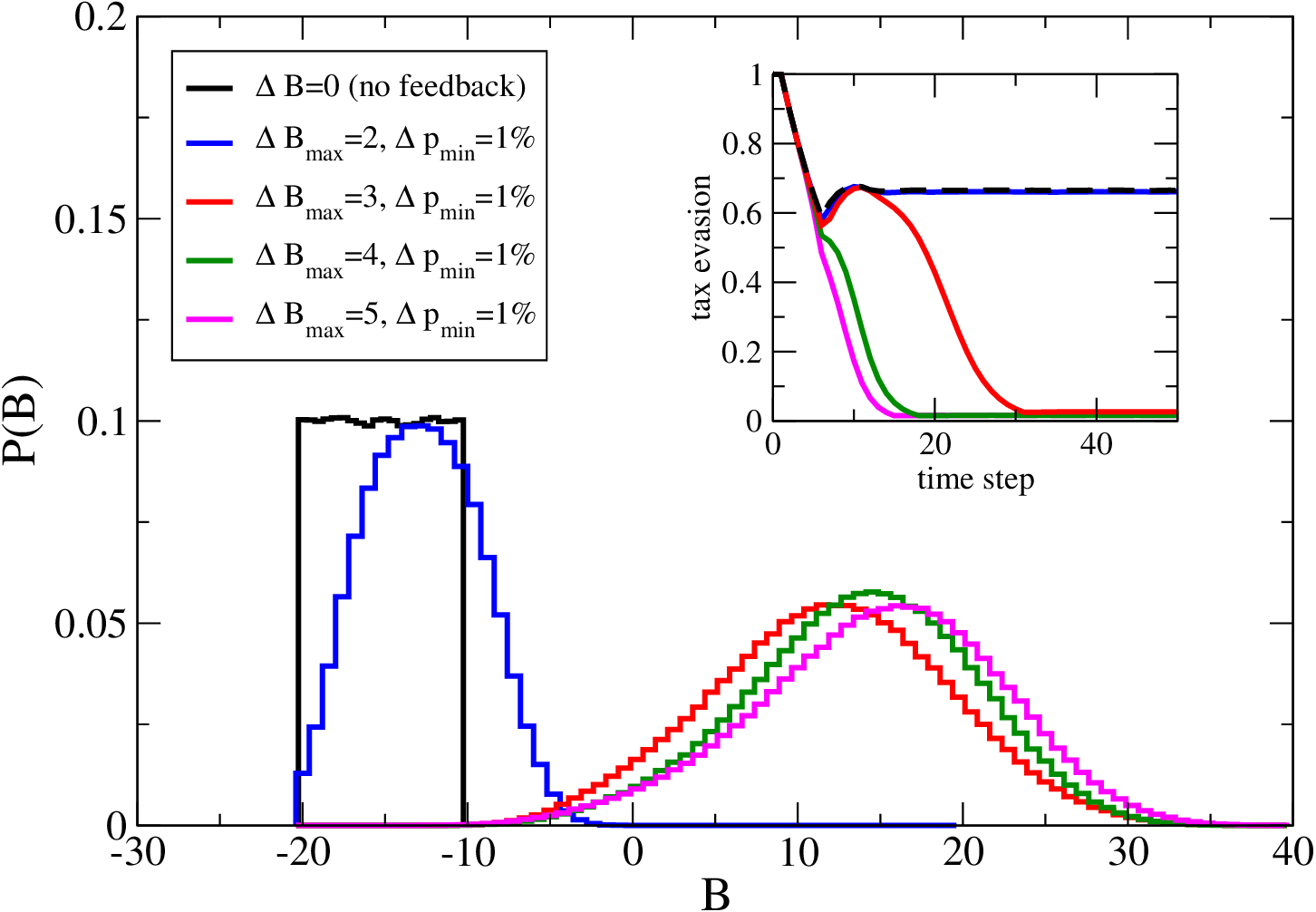}
\includegraphics[width=12cm,clip=true]{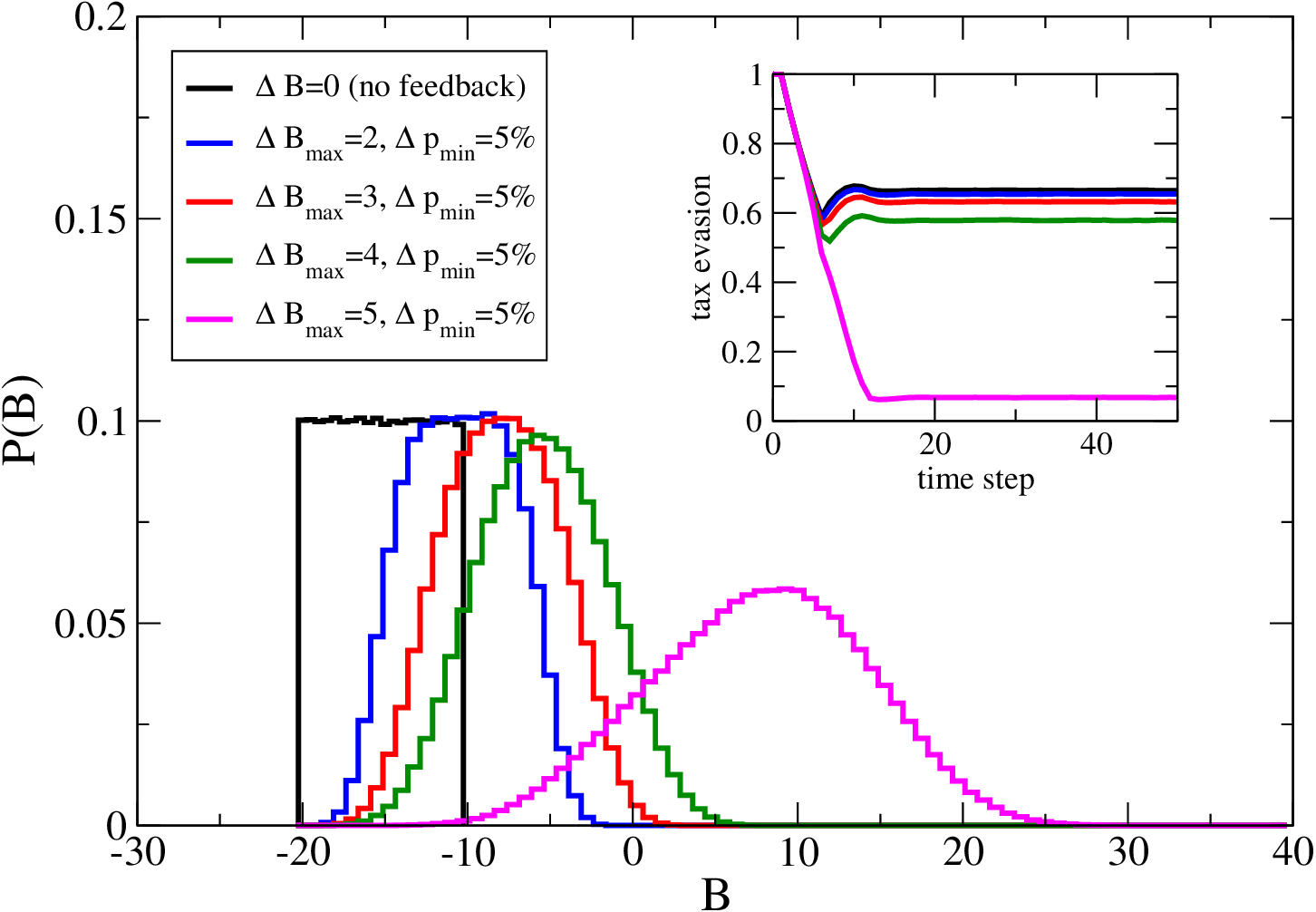}
\end{center}
\caption{Main panel: The distribution of local fields for 
various rates of adaption $\Delta B_{max}$. The case $\Delta B_{max}=0$
corresponds to the case with no feedback of public goods which is identical
to the distribution used in Fig. \ref{fig0}a.
Results are reported for a societcy consisting of $100\%$ a-types, 
enforcement with fixed compliance period $h=5$ 
and audit probability $p_a=10\%$. The initial distribution of local fields
is confined to the range $-20 < B_i < -10$.
Inset: Time evolution of tax evasion (fraction of non-compliant
agents $p_{non-cp}$).
Top panel: Threshold of perception $\Delta p_{min}=1\%$, Lower panel: Threshold
of perception $\Delta p_{min}=5\%$.}
\label{fig2}
\end{figure}

In Fig. \ref{fig2} we investigate the influence of public goods provision on a 
society consisting of $100\%$ selfish a-type agents for various rates of 
adaption 
$\Delta B_{max}$ and two thresholds of perception 
$\Delta p_{min}=1\%$ (upper panel) and $\Delta p_{min}=5\%$ (lower panel).  
This feedback influences on the local field distribution
$P(B_i)$ which is now determined self-consistently as described above.
At time step $t_n=0$ selfish a-types are given a flat distribution of 
local fields confined to the range
$-10 < B_i < -20$ and which is shown by the black curve in the main panels of
Fig. \ref{fig2}. Note that this is a generalization of \cite{lz08} which
have used a single field value.
However, the feedback of public goods leads to a redistribution
of $P(B_i)$ which for large times becomes stationary and for different
$\Delta B_{max}$ and $\Delta p_{min}$ is shown in the main panel of
Fig. \ref{fig2}.

Interestingly one observes two regimes in the distribution
of local fields. For small $\Delta B_{max}$ 
the distribution transforms into a Gaussian and shifts to larger mean values
$B_{mean}$ which are still in the range $B_{mean}<0$, i.e. the majority of
agents is still characterized as non-compliant taxpayers. 
From the insets to Fig. \ref{fig2} we observe that the dynamics in this
case is similar to the one without feedback of public goods. Due to the
shift of $P(B)$ to larger mean values a slight reduction in the stationary
value of tax evasion becomes visible, especially for
larger thresholds of perception.

Above a critical value of the adaption rate
($\Delta B_{max}^{crit}\approx 2.5$ for $\Delta p_{min}=1$ and $\Delta B_{max}^{crit}\approx 4.5 $ for $\Delta p_{min}=5$) one observes a broad distribution
with a mean value which is now deep in the positive $B$-regime, i.e. most
of the selfish a-type agents show a dominant compliant behavior. 
As a consequence
the stationary value of tax evasion (insets to Fig. \ref{fig2})
is now strongly reduced and the remaining tax evaders are
solely due to the 'negative $B$' agents in the tail of $P(B)$.

We emphasize that the distributions shown in the main panels
of Fig. \ref{fig2} are stationary distributions, i.e. obtained after
$t=200$ time steps. We have checked that the same distributions are
obtained after $t=5,000$ time steps in both regimes.
Clearly, the crossover between the two regimes is
due to a self-sustaining effect when the agents start to become
compliant and thus enhance the provision of public goods,
combined with the audit which enforces the detected agent to change
its behavior from $S_i = -1$ to $S_i=+1$ over the following $h$ periods.
According to the mechanism  summarized in 
the third line of Tab. \ref{tab1} a regime where the provision of public
goods has significantly increased, induces a behavioral change of 
non-compliant agents upon auditing. This in turn enhances
the provision of public goods and is responsible for the self-sustaining
effect leading to the crossover between the two regimes.

An interesting (but also pathological) situation
may arise when the distribution of local fields is centered
around $B=0$. In this case there may appear long term oscillations 
between both regimes, i.e. between substantial and small fractions
of tax evaders. For the present parameters such situation occurs for 
threshold value $\Delta p_{min}=0$ and a critical adaption rate
of $\Delta B_{max}=1$. For values $\Delta B_{max}>1$ the resulting
distributions are similar to the case $\Delta p_{min}=1\%$ but shifted
to larger (positive) mean values.

\begin{figure}[htb]
\begin{center}
\includegraphics[width=12cm,clip=true]{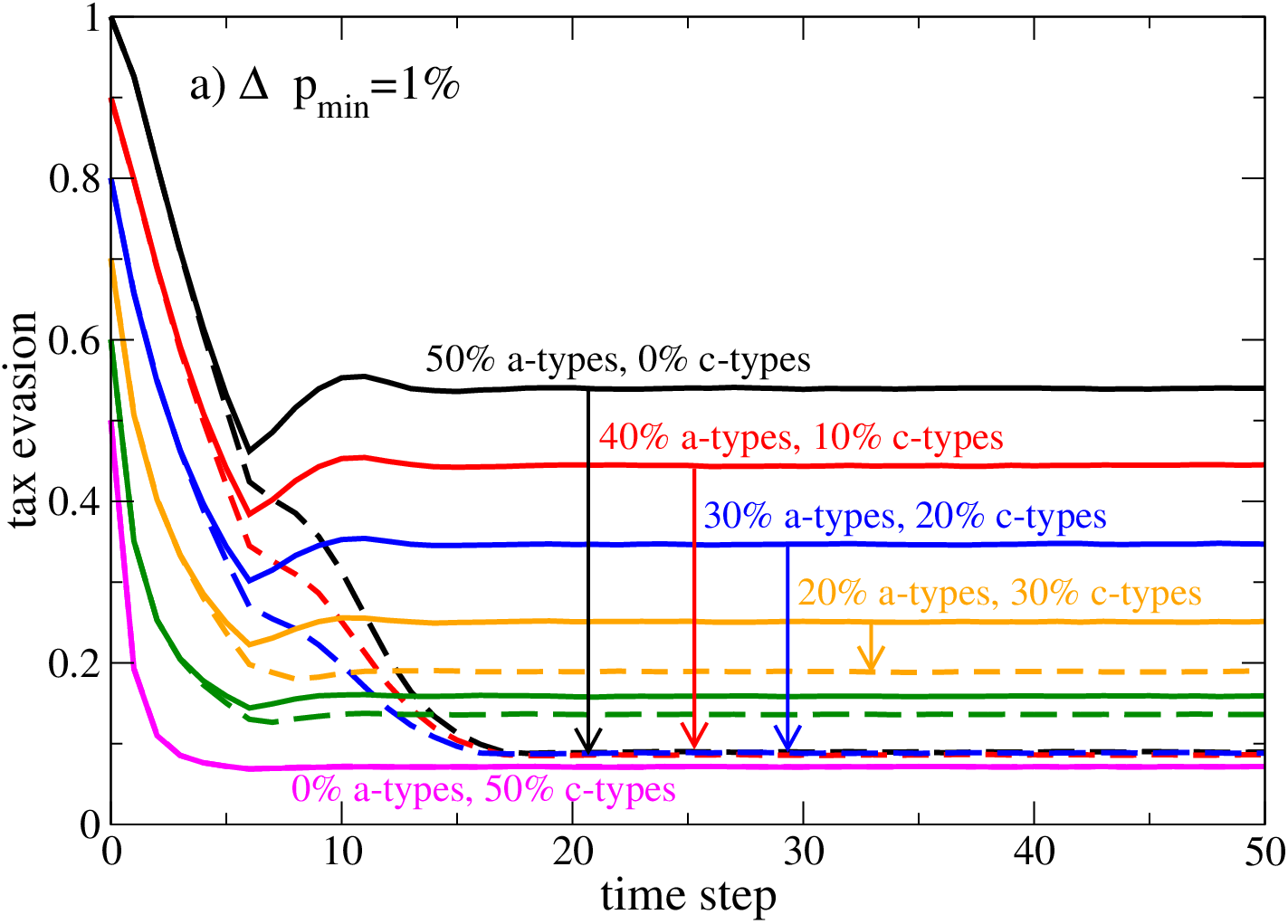}
\includegraphics[width=12cm,clip=true]{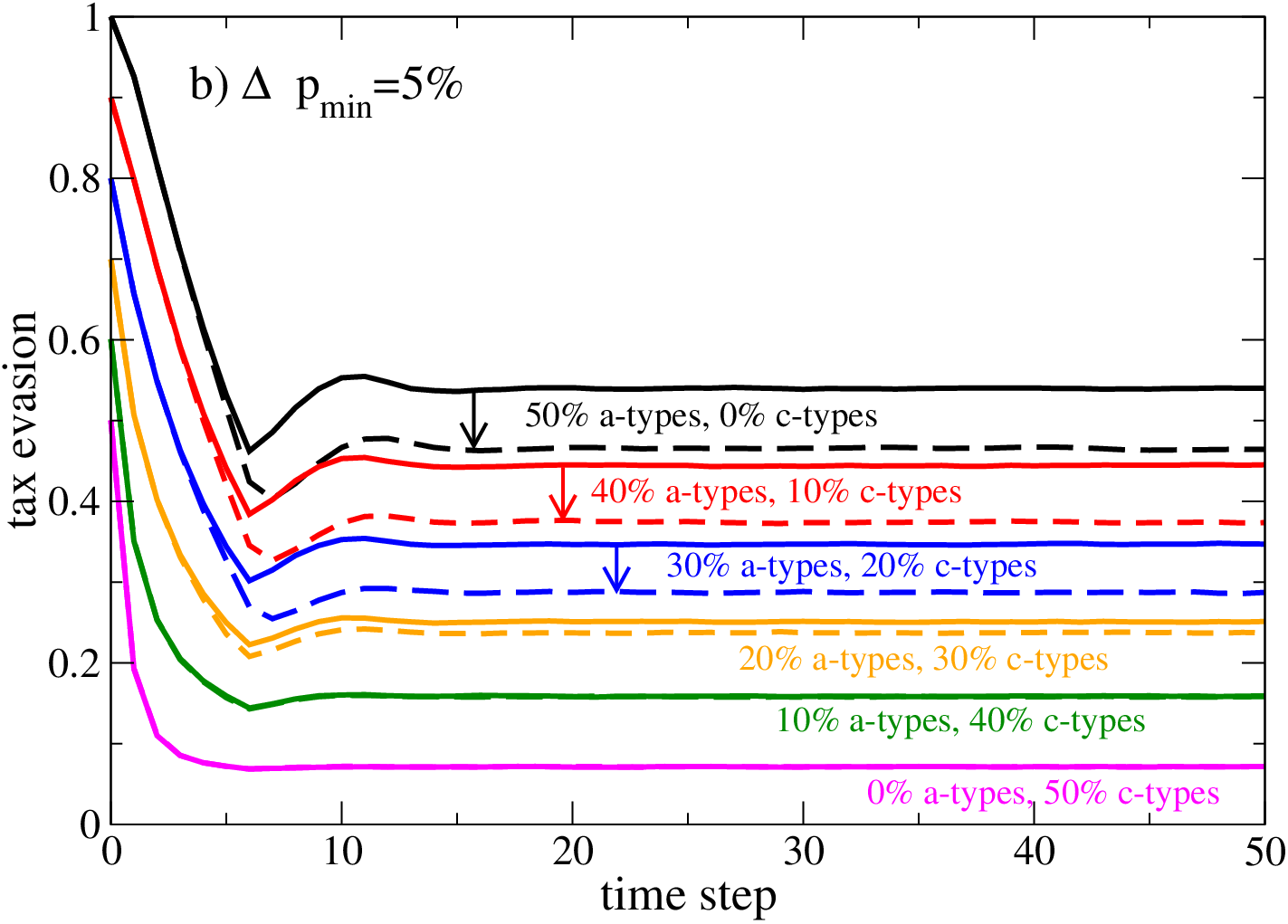}
\end{center}
\caption{Time evolution of tax evasion (fraction of non-compliant
agents $p_{non-cp}$) for a society consisting of $35\%$ b-types
and $15\%$ d-types. The fraction of a- and c-types is indicated
in the panels. Solid lines report the result without and dashed lines
the result including the feedback of public goods provision.
Results are obtained for enforcement with fixed compliance period $h=5$ 
and audit probability $p_a=10\%$. The initial distribution of local fields
is confined to the range $-20 < B_i < -10$ and the rate of adaption is
$\Delta B_{max}=4$. 
Panel (a): Threshold of perception $\Delta p_{min}=1\%$, Panel (b): Threshold
of perception $\Delta p_{min}=5\%$.}
\label{fig3}
\end{figure}

How are the above results generalized within
a multi-agent model? To answer this question we calculate the
fraction of tax evasion for a society consisting of $35\%$ b-types,
$15\%$ d-types and the residual share of a- and c-types is
varied. Note that the percentage of $15\%$ d-types can be 
motivated from estimates for U.S. households taken from
\citet{and98}. Similar agent shares have also been investigated 
by \citet{hopi10} and \citet{pick12}. We choose an adaption
rate of $\Delta B_{max}=4$ so that for small threshold of perception
$p_{min}=1\%$ (panel a) the majority of a-type agents undergoes
a transition towards compliant behavior (cf. Fig. \ref{fig2}).
As a consequence the fraction of evaders for societies consisting
of $50\%$, $40\%$, and $30\%$ a-types is drastically reduced as
is apparent from Fig. \ref{fig3}a.
This reduction becomes less pronounced for smaller a-type
shares (e.g. $20\%$ a-types).  In fact, the transition of a-types 
from non-compliance to compliance observed
in Fig. \ref{fig2} is driven by a significant change (enhancement) in the
public goods provision. Now a large share of c-types obviously constitutes
the fraction of compliant agents and therefore to the public goods provision 
but this contribution is constant over time. On the
other hand, the behavioral change of a-types is induced by a concomitant
change in the public goods provision and this change is naturally
reduced when the fraction of c-types becomes to large.
Therefore we observe only moderate to small reduction in 
tax evasion for a-type shares of $20\%$, $10\%$, and $0\%$.

Fig. \ref{fig3}b reports analogous 
results for a larger threshold of perception $p_{min}=5\%$.
From the previous results shown in Fig. \ref{fig2} we know that for
this parameter the mean value of the field distribution for
a-types is still in the negative range so that non-compliance is predominant. 
As a consequence also the
effective share of non-compliant agents for the composed society is only
moderately reduced  upon allowing for the feedback from public goods
provision.

\section{Discussion and Outlook}\label{sec4}
We have investigated the influence of public goods provision
on tax evasion within an econophysics model
which describes the interaction of various behavioral types of agents.
Within our framework only selfish (a-types) agents are susceptible
to the provision of public goods and we have shown
that their behavioral attitude towards tax evasion 
is determined by a dynamic local 'field-parameter' $B_i$.
As a result we find that an initially flat distribution
$P(B_i)$ is altered self-consistently due to the feedback
of public goods provision leading to a generic shift towards
larger field value (increased 'morality' of agents). This in
turn can reduce tax evasion in agreement
with previous investigations within the economics domain \cite{szabo10}.
Also experimental data in the context of tax evasion \cite{alm1,alm2}
support
the finding that compliance increases when taxpayers receive a
public good in return for their payment.
On the other hand \cite{ho2014} implements public goods provision
via an utility function which is the crucial factor for
the counterintuitive result that income tax evasion may increase
providing a higher level of public goods. With regard to our
present approach the different result can be traced back to
a different behavior upon the increase of public goods provision
(cf. third line of Tab. \ref{tab1}) where the selfish
agent in  \cite{ho2014} does not increase its moral attitude.

Two parameters govern the feedback of 
public goods provision to the behavioral change of selfish agents:
(i) the rate of adaption $\Delta B_{max}$ and (ii) the threshold
of perception $\Delta p_{min}$ towards changes in public goods provision.
$\Delta B_{max}$ is an endogenous parameter of the agents
which may depend on age but otherwise is not susceptible to external
influence. On the other hand, policymakers can to a certain extend 
influence on the threshold of perception $\Delta p_{min}$ by
means of targeted public relations. In practice this may be achieved by 
announcements
via public media also about minor state projects.

Since often public goods provision is more important
for the elder generation (as e.g. in case of health care)
it may also partially account for their stronger tax morale
as found by \citet{nordblom12}.
However, in our econophysics model we have assumed a constant 
(over time) distribution
of agent types. On the other hand, \citet{nordblom12} have
set up an economic model which describes how social beings 
update their personal norms.
Implementing these mechanisms in our model would allow for the
transformation between different agent types and is an interesting
perspective for future research.

\section*{Acknowledgements}

We would like to thank Robert Axtell, C\'{e}cile Bazart, Wolfram Berger, 
Toni Ll\'{a}cer, and two anonymous referees for helpful 
comments and suggestions. 
However, errors remain our own. 
Sascha Hokamp owes special thanks to Janina Schnieders, Petra Lackinger, 
Julia Pistier, Faris Al-Mashat, and Sebastian L\"uke. Financial support for 
Sascha Hokamp from the European Social Simulation Association and the German 
Academic Exchange Service is gratefully acknowledged.


\begin{thebibliography}{}
\bibitem[\protect\citeauthoryear{Allingham and Sandmo}{1972}]{alling72}
\textsc{Allingham, M. G.} and \textsc{A. Sandmo} (1972).       
Income Tax Evasion: A Theoretical Analysis.                            
\textit{Journal of Public Economics} \textbf{1}, 323--338. 

\bibitem[\protect\citeauthoryear{Alm et al.}{1992a}]{alm1}\textsc{Alm, J.}
and \textsc{B. Jackson} and \textsc{M. McKee} 
(1992a). Estimating the Determinants of Taxpayer Compliance with Experimental
Data. \textit{National Tax Journal} \textbf{45}(1), 107 -- 114.   

\bibitem[\protect\citeauthoryear{Alm et al.}{1992b}]{alm2}\textsc{Alm, J.}
and \textsc{G. H. McClelland} and \textsc{W. D. Schulze} 
(1992b). Why do people pay taxes? \textit{Journal of Public Economics} 
\textbf{48}, 21 -- 38.  

\bibitem[\protect\citeauthoryear{Alm}{2010}]{ALM2010}
\textsc{Alm, J.} (2010). Testing Behavioral Public Economic Theories in the Laboratory. 
\textit{National Tax Journal} \textbf{63} (4), 635 -- 658.


\bibitem[\protect\citeauthoryear{Andrei et al.}{2014}]{andr13}
\textsc{Andrei, A.}, \textsc{K. Comer} and \textsc{M. Koehler} 
(2014). An agent-based model of network effects on tax compliance and evasion. 
\textit{Journal of Economic Psychology} \textbf{40}, 119 -- 133.  

\bibitem[\protect\citeauthoryear{Andreoni et al.}{1998}]{and98}
\textsc{Andreoni, J.}, \textsc{B. Erard} and \textsc{J. Feinstein} 
(1998). Tax Compliance. \textit{Journal of Economic Literature} \textbf{36}(2), 
818--860.                                                                      

\bibitem[\protect\citeauthoryear{Antunes et al.}{2007}]{antun07}
\textsc{Antunes, L.}, \textsc{Jo. Balsa}, \textsc{A. Respicio} and 
\textsc{H. Coelho} (2007). Tactical Exploration of Tax Compliance  
Decisions in Multi-agent Based Simulation, in: Luis Antunes and Keiki Takadama 
(Eds.), \textit{Multi-Agent-Based Simulation VII}, International               
Workshop MABS 2006, LNAI 4442, Heidelberg: Springer, 80--95.                           

\bibitem[\protect\citeauthoryear{Bazart and Pickhardt}{2011}]{bapick11}
\textsc{Bazart, C.} and \textsc{M. Pickhardt} 
(2011). Fighting Income Tax Evasion with Positive Rewards. 
\textit{Public Finance Review} \textbf{39}(1), 
124--149.                                                                      


\bibitem[\protect\citeauthoryear{Bloomquist}{2004}]{bloom04}
\textsc{Bloomquist, K. M.} (2004). Modeling Taxpayers' Response to 
Compliance Improvement Alternatives. Paper presented at the Annual Conference 
of the North American Association for Computational Social and Organizational 
Sciences, Pittsburgh, PA.                                                     
                                                                              
\bibitem[\protect\citeauthoryear{Bloomquist}{2006}]{bloom06}                  
\textsc{Bloomquist, K. M.} (2006). A Comparison of Agent-based Models of     
Income Tax Evasion. \textit{Social Science Computer Review} \textbf{24}(4),   
411--425.                                                                     

\bibitem[\protect\citeauthoryear{Bloomquist}{2008}]{bloom08}                  
\textsc{Bloomquist, K. M.} (2008). Taxpayer Compliance Simulation: A Multi-Agent Based Approach, In: Edmonds, B., Hern{\'a}ndez, C. and Troitzsch, K. G. (eds.) \textit{Social Simulation: Technologies, Advances and New Discoveries}. Premier References Series, Chap. 2, 13--25.                                            

\bibitem[\protect\citeauthoryear{Bloomquist}{2011}]{BLOO4}                  
\textsc{Bloomquist, K. M.} (2011). 
Tax Compliance as an Evolutionary Coordination Game: An Agentbased Approach. 
\textit{Public Finance Review} \textbf{39} (1), 25 -- 49.
                         
\bibitem[\protect\citeauthoryear{Bordignon}{1993}]{BORD}                  
\textsc{Bordignon, M.} (1993). 
A Fairness Approach to Income Tax Evasion.
\textit{Journal of Public Economics} \textbf{52}, 345 -- 362.

 

\bibitem[\protect\citeauthoryear{Cowell and Gordon}{1988}]{COGO}      
\textsc{Cowell, F. A.} and \textsc{J. P. F. Gordon}(1988).
Unwillingness to Pay - Tax Evasion and Public Goods Provision. 
\textit{Journal of Public Economics} \textbf{36}, 305 -- 321.

\bibitem[\protect\citeauthoryear{Cowell}{1992}]{COWE}      
\textsc{Cowell, F. A.} (1992).
Tax Evasion and Inequity. \textit{Journal of Economic Psychology}
\textbf{13}, 521 -- 543.     

\bibitem[\protect\citeauthoryear{Davis et al.}{2003}]{dav03}      
\textsc{Davis, J. S.}, \textsc{G. Hecht} and \textsc{J. D. Perkins} 
(2003). Social Behaviors, Enforcement, and Tax Compliance Dynamics.     
\textit{The Accounting Review} \textbf{78}(1), 39--69.                  


\bibitem[\protect\citeauthoryear{Falkinger}{1988}]{FALK1}      
\textsc{Falkinger, J.} (1988)
Tax Evasion and Equity: A Theoretical Analysis. 
\textit{Public Finance/Finances Publiques} \textbf{43}, 388 -- 395.

\bibitem[\protect\citeauthoryear{Falkinger}{1991}]{FALK2}      
\textsc{Falkinger, J.} (1991)
On Optimal Public Good Provision with Tax Evasion. 
\textit{Journal of Public Economics} \textbf{45}, 127 -- 133.

\bibitem[\protect\citeauthoryear{Falkinger and Walter}{1991}]{FAWA}      
\textsc{Falkinger, J.} and \textsc{H. Walther} (1991).
Rewards versus Penalties: On a New Policy against Tax Evasion. 
\textit{Public Finance Quarterly} \textbf{19}, 67 -- 79.

\bibitem[\protect\citeauthoryear{Falkinger}{1995}]{falki}      
\textsc{Falkinger, J.} 
(1995). Tax Evasion, Consumption of Public Goods and Fairness.     
\textit{Journal of Economic Psychology} \textbf{16}, 63--72.                  
        
\bibitem[\protect\citeauthoryear{Heinemann and Schneider}{2011}]{heinemann12}
\textsc{Heinemann, F.} and \textsc{F. Schneider} (2011). 
Religion and the Shadow Economy. \textit{ZEW Discussion Papers} \textbf{11-038}.
    
\bibitem[\protect\citeauthoryear{Hokamp and Pickhardt}{2010}]{hopi10}
\textsc{Hokamp, S.} and \textsc{M. Pickhardt} (2010). Income Tax 
Evasion in a Society of Heterogeneous Agents - Evidence from an Agent-based 
Model. \textit{International Economic Journal} \textbf{24}(4), 541--553.    

\bibitem[\protect\citeauthoryear{Hokamp}{2013}]{ho2013}
\textsc{Hokamp, S.} (2013). Income Tax Evasion and Public Goods Provision -- Theoretical Aspects and Agent-based Simulations
\textit{Brandenburg University of Technology Cottbus, Ph.D. Thesis}.

\bibitem[\protect\citeauthoryear{Hokamp}{2014}]{ho2014}
\textsc{Hokamp, S.} (2014). Dynamics of tax evasion with back auditing, social norm updating and public goods provision - An agent-based simulation. 
\textit{Journal of Economic Psychology} \textbf{40}, 187 -- 199.

\bibitem[\protect\citeauthoryear{Hokamp and Seibold}{2014}]{hoseib}
\textsc{Hokamp, S.} and \textsc{G. Seibold} (2014). How much Rationality tolerates the Shadow Economy? - An Agent-based Econophysics Approach. 
\textit{Advances in Social Simulation; Conference Series: Advances in Intelligent Systems and Computing} \textbf{229}, 119--128.

\bibitem[\protect\citeauthoryear{Ising}{1925}]{ising25}\textsc{Ising, E.}
 (1925). Beitrag zur Theorie des Ferromagnetismus. 
\textit{Zeitschrift f\"ur Physik} \textbf{31}(1), 253--258.

\bibitem[\protect\citeauthoryear{Kirchler}{2007}]{KIRC}
\textsc{Kirchler, E.} (2007).
The Economic Psychology of Tax Behavior. \textit{Cambridge University Press, 
Cambridge}.

\bibitem[\protect\citeauthoryear{Korobow et al.}{2007}]{koro07}\textsc{Korobow,
 A.}, \textsc{C. Johnson} and \textsc{R. Axtell}  (2007).             
An Agent-based Model of Tax Compliance with Social Networks. \textit{National  
Tax Journal} \textbf{60}(3), 589-610.                                          

\bibitem[\protect\citeauthoryear{Krauth}{2006}]{krauth}\textsc{Krauth, W.} 
(2006). \textit{Statistical Mechanics; Algorithms and Computations}. Oxford    
University Press.                                                              

\bibitem[\protect\citeauthoryear{Lima and Zaklan}{2008}]{lz08}\textsc{Lima, 
F. W. S.} and \textsc{G. Zaklan} (2008). A Multi-agent-based      
Approach to Tax Morale. \textit{International Journal of Modern Physics C:  
Computational Physics and Physical Computation} \textbf{19}(12), 1797--1808.

\bibitem[\protect\citeauthoryear{Lima}{2010}]{lima10}\textsc{Lima, F. 
W. S.} (2010). Analysing and Controlling the Tax Evasion Dynamics via        
Majority-Vote Model. \textit{Journal of Physics: Conference Series}          
\textbf{246}, 1--12.                                                         

\bibitem[\protect\citeauthoryear{M\'{e}der et al.}{2012}]{meder10}
\textsc{M\'{e}der, Z. Z.}, \textsc{A. Simonovits} and 
\textsc{J. Vincze} (2012). Tax Morale and Tax Evasion: Social 
Preferences and Bounded Rationality. \textit{Economic Analysis and Policy} 
\textbf{42}(2), 171 --188.                                                    

\bibitem[\protect\citeauthoryear{Mittone and Patelli}{2000}]{mp00}
\textsc{Mittone, L.} and \textsc{P. Patelli} (2000). Imitative Behaviour 
in Tax Evasion, in: Francisco Luna and Benedikt Stefansson (eds.),             
\textit{Economic Simulations in Swarm: Agent-based Modelling and Object        
Oriented Programming}, Dordrecht, Boston, London: Kluwer Academic Publishers,  
133--158. 

\bibitem[\protect\citeauthoryear{Mittone}{2006}]{mittone06}
\textsc{Mittone, L.} (2006). Dynamic Behaviour in Tax Evasion: An Experimental Approach. 
\textit{The Journal of Socio-Economics} \textbf{35}, 813 -- 835. 

\bibitem[\protect\citeauthoryear{Musgrave}{1939}]{MUSG1}
\textsc{Musgrave, R. A.} (1939).
The Voluntary Exchange Theory of Public Economy. 
\textit{The Quarterly Journal of Economics} \textbf{53}, 213 -- 237.

\bibitem[\protect\citeauthoryear{Musgrave}{1969}]{MUSG2}
\textsc{Musgrave, R. A.} (1969).
Provision for Social Goods. 
\textit{Public Economics, McMillan, London}, 124 -- 144.

\bibitem[\protect\citeauthoryear{Musgrave}{1999}]{MUSG3}
\textsc{Musgrave, R. A.} (1999).
The Nature of the Fiscal State: The Roots of My Thinking. 
\textit{Public Finance and Public Choice, Cambridge: MIT Press}, 
29 -- 49.

\bibitem[\protect\citeauthoryear{Nordblom and \v{Z}amac}{2012}]{nordblom12}
\textsc{Nordblom, K.} and \textsc{J. \v{Z}amac} (2012). Endogenous 
Norm Formation Over the Life Cycle - The Case of Tax Morale. 
\textit{Economic Analyis \& Policy} \textbf{42}(2), 153--170.    


\bibitem[\protect\citeauthoryear{Pellizzari and Rizzi}{2014}]{pelli13}
\textsc{Pellizzari, P.} and \textsc{D. Rizzi} 
(2014). Citizenship and power in an agent-based model of tax compliance
with public expenditure. \textit{Journal of Economic Psychology}
\textbf{40}, 35 -- 48.

\bibitem[\protect\citeauthoryear{Pickhardt}{2003}]{PICK1}
\textsc{Pickhardt, M.} (2003).
Studien zur Theorie \"offentlicher G\"uter. 
\textit{Metropolis, Marburg, Germany}.

\bibitem[\protect\citeauthoryear{Pickhardt}{2006}]{PICK3}
\textsc{Pickhardt, M.} (2006).
Fifty Years after Samuelson's `The Pure Theory of Public Expenditure`: What Are We Left with? \textit{Journal of the History of Economic Thought} 
\textbf{28} (4) 439 -- 460.

\bibitem[\protect\citeauthoryear{Pickhardt and Prinz}{2014}]{pickprinz}
\textsc{Pickhardt, M.} and \textsc{A. Prinz} (2014). Behavioral Dynamics
of Tax Evasion - A Survey.
\textit{Journal of Economic Psychology} \textbf{40}, 1 -- 19.

\bibitem[\protect\citeauthoryear{Pickhardt and Seibold}{2014}]{pick12}
\textsc{Pickhardt, M.} and \textsc{G. Seibold} (2014). Income Tax Evasion Dynamcis: Evidence from an Agent-based Econophysics Model.
\textit{Journal of Economic Psychology} \textbf{40}, 147 -- 160.

\bibitem[\protect\citeauthoryear{Samuelson}{1954}]{SAMU1}
\textsc{Samuelson, P. A.}(1954).
The Pure Theory of Public Expenditure.
\textit{The Review of Economics and Statistics} \textbf{36}, 387 -- 389.

\bibitem[\protect\citeauthoryear{Samuelson}{1955}]{SAMU2}
\textsc{Samuelson, P. A.}(1955).
Diagrammatic Exposition of a Theory of Public Expenditure. 
\textit{Review of Economics and Statistics} \textbf{37}, 350 -- 356.

\bibitem[\protect\citeauthoryear{Samuelson}{1969}]{SAMU3}
\textsc{Samuelson, P. A.}(1969).
Pure Theory of Public Expenditure. 
\textit{Public Economics, McMillan, London}, 98 -- 123.

\bibitem[\protect\citeauthoryear{Samuelson and Nordhaus}{1998}]{SANO}
\textsc{Samuelson, P. A.} and \textsc{W. D. Nordhaus} (1998).
Economics. \textit{McGraw-Hill, New York, 16th edition}.

\bibitem[\protect\citeauthoryear{Seibold and Pickhardt}{2013}]{seibpick12}
\textsc{Seibold,                                                          
G.} and \textsc{M. Pickhardt} (2013). Lapse of time effects on tax evasion in an agent-based Econophysics model. \textit{Physica A} \textbf{392}(9),
2079--2087.

\bibitem[\protect\citeauthoryear{Szab\'{o} et al.}{2009}]{szabo09}
\textsc{Szab\'{o}, A.}, \textsc{L. Guly\'{a}s} and 
\textsc{I. J. T\'{o}th} (2009). Sensitivity Analysis of a Tax Evasion 
Model Applying Automated Design of Experiments, in: Luis S. Lopes, Nuno Lau,  
Pedro Mariano, and Luis M. Rocha (Eds.), \textit{Progress in Artificial       
Intelligence}, 14th Portuguese Conference on Artificial Intelligence,         
EPIA 2009, Aveiro, Portugal, LNAI 5816, Berlin Heidelberg: Springer, 572-583. 

\bibitem[\protect\citeauthoryear{Szab\'{o} et al.}{2010}]{szabo10}
\textsc{Szab\'{o}, A.} and \textsc{L. Guly\'{a}s} and 
\textsc{I. J. T\'{o}th}  (2010). Simulating Tax Evasion with Utilitarian
Agents and Social Feedback. \textit{International Journal of Agent Technologies and Systems} \textbf{2}(1), 16 -- 30. 

\bibitem[\protect\citeauthoryear{Zaklan et al.}{2008}]{zak08}\textsc{Zaklan,
G.}, \textsc{F. W.S. Lima} and \textsc{F. Westerhoff} (2008).
Controlling Tax Evasion Fluctuations. \textit{Physica A: Statistical
Mechanics and its Applications} \textbf{387}(23), 5857--5861.

\bibitem[\protect\citeauthoryear{Zaklan et al.}{2009}]{zak09}\textsc{Zaklan,
G.}, \textsc{F. Westerhoff} and \textsc{D. Stauffer} (2009).
Analysing Tax Evasion Dynamics via the Ising Model. \textit{Journal of
Economic Interaction and Coordination} \textbf{4}, 1--14.


\end{thebibliography}
\end{document}